\documentclass[12pt]{iopart}

\usepackage{iopams}

\usepackage{verbatim,dsfont}
\usepackage{graphicx}

\newcommand{\be}{\begin{equation}}
\newcommand{\ee}{\end{equation}}
\newcommand{\ba}{\begin{array}}
\newcommand{\ea}{\end{array}}
\newcommand{\bea}{\begin{eqnarray}}
\newcommand{\eea}{\end{eqnarray}}

\newcommand{\LL}{\mathbb L}
\newcommand{\calA}{{\cal A }}
\newcommand{\calC}{{\cal C }}
\newcommand{\calL}{{\cal L }}
\newcommand{\calH}{{\cal H }}

\newcommand{\calE}{{\cal E }}

\newcommand{\calZ}{{\cal Z }}
\newcommand{\calP}{{\cal P }}

\newcommand{\calS}{{\cal S }}
\newcommand{\calG}{{\cal G }}

\newcommand{\ZZ}{\mathbb{Z}}
\newcommand{\CC}{\mathbb{C}}

\newcommand{\la}{\langle}
\newcommand{\ra}{\rangle}

\newcommand{\nn}{\nonumber}
\newcommand{\rank}{\mathop{\mathrm{rank}}\nolimits}

\newcommand{\trace}{\mathop{\mathrm{Tr}}\nolimits}

\newtheorem{dfn}{Definition}
\newtheorem{lemma}{Lemma}
\newtheorem{prop}{Proposition}
\newtheorem{theorem}{Theorem}

\begin{document}

\title[Thermodynamic stability criteria for a quantum memory]{Thermodynamic stability criteria for a quantum memory based on stabilizer and subsystem codes}

\author{Stefano Chesi and Daniel Loss}

\address{Department of Physics, University of Basel, Klingelbergstrasse 82, 4056 Basel, Switzerland}
\author{Sergey Bravyi and Barbara M. Terhal}
\address{IBM Watson Research Center, Yorktown Heights, NY 10598,
USA}
\begin{abstract}
We discuss and review several thermodynamic criteria that have been
introduced to characterize the thermal stability of a
self-correcting quantum memory. We first examine the use of
symmetry-breaking fields in analyzing the properties of
self-correcting quantum memories in the thermodynamic limit:
we show that the thermal expectation values of all logical operators
vanish for any stabilizer and any subsystem code in any spatial
dimension.
On the positive side, we generalize the results in [R.~Alicki {\it
et al.}, arXiv:0811.0033.] to obtain a general upper bound on the
relaxation rate of a quantum memory at nonzero temperature, assuming
that the quantum memory interacts via a Markovian master equation
with a thermal bath. This upper bound is applicable to quantum
memories based on either stabilizer or subsystem codes.
\end{abstract}

\submitto{\NJP Special Issue on ``Quantum Information and Many-Body
Theory"}

\maketitle

\section{Introduction}

Thermal fluctuations pose a serious problem for reliable, passive,
information storage since any open system eventually reaches a
thermal equilibrium state in which all encoded information is lost.
Fortunately, it has been shown that quantum information can be
reliably stored for arbitrary long times in, say, a 2D quantum
memory \cite{Dennis2002} by means of active error correction and
entropy removal. However, the implementation of active error
correction implying extensive and fast classical input/output to the
quantum memory poses a serious (but hopefully not insurmountable)
experimental challenge.

The central idea behind self-correcting classical or quantum
memories is to do without active error-correction and prevent
thermalization and build-up of entropy by the presence of
macroscopic ``energy barriers" separating encoded states.

The idea of such self-correcting quantum memory was first introduced
in \cite{Dennis2002} and can be viewed as an extension of the ideas
of topological protection developed by Kitaev \cite{Kitaev1997}. In
\cite{Dennis2002} it was argued that the 2D surface or toric code
(2D Kitaev model) would not be a self-correcting memory, but a 4D
surface code (4D Kitaev model) generalization was presented which
would be thermally stable. Unfortunately, three spatial dimensions
is all the room that the natural world seems to provide.

It is thus of interest to (1) either come up with models for
self-correcting quantum memories in 3 or fewer dimensions,
or (2) show that low-dimensional quantum physics does not
allow for such passive stability. The latter possibility would
provide evidence that genuine quantum phases of nature, such as
topological phases, would be confined to the domain of finite
systems and low temperatures: in the thermodynamic limit thermal
fluctuations would destroy the quantum order at any nonzero
temperature. Such a no-go possibility would also lend support to the
thought that macroscopic quantum states suffer from intrinsic
decoherence (see \cite{Alicki:pm} for thoughts in this direction).
In this sense we believe that the question of thermal stability of a
passive quantum memory is one of fundamental interest.

In fact, the thermal stability question can also be viewed as a
question about the nature of the excitations of the quantum memory
model. For 2D topological models these excitations are point-like
pairs of anyons. If we paraphrase the macroscopic energy barrier
requirement of \cite{Bravyi2008} in terms of the nature of
excitations, it relates to a condition that the elementary
excitations are extended objects; they are the boundary of a two or
higher-dimensional surface.

In \cite{Bacon2005} the subject of self-correcting quantum memories
was brought to the fore. Bacon introduced two models, now called the
2D Bacon-Shor code or quantum compass model, and the 3D Bacon-Shor
code, which are examples of quantum subsystem codes. The 3D
Bacon-Shor model may or may not be an example of a self-correcting
quantum memory; it is an open question how to analyze its thermal
stability.

The analysis of the thermal stability of a quantum stabilizer or
quantum subsystem code model in a thermodynamic sense is the subject
of this paper. Let us discuss some of the literature on this
subject.

Necessary criteria for thermal stability of a quantum memory were
formulated in \cite{Bravyi2008} (see also \cite{Kay2008}) in terms
of a macroscopic distance of the underlying quantum code (i.e. zero
temperature topological order) and the presence of macroscopic
energy barriers. It was shown in that paper that all 1D and 2D local
stabilizer codes fail to meet these criteria. The advantage of this
approach is that it allows for very general no-go results. A
disadvantage is that it does not make contact with any operational
or thermodynamic expression of thermal stability. In particular, to
prove positive results on particular quantum memory models, it is
necessary to more thoroughly analyze the thermodynamics of an open
quantum memory.
The intuition that underlies the idea of a self-correcting quantum
memory is that errors of increasing weight should map encoded states
onto excited states with increasingly higher energy. If the quantum
code has a macroscopic distance which scales with system-size, then
high-weight errors will have to happen in order to map one encoded
state onto another. But such high-weight errors will, if the memory
is self-correcting, correspond to high-energy states, hence there
would be (macroscopic) energy barriers between different encoded
states. In the second part of our paper, Section \ref{sec:use} we
will indeed see that the energy associated with high-weight errors
corresponding to so-called bad syndromes, will play a crucial role
in bounding the quantum memory relaxation rate.

Specific results ruling out the existence of finite temperature
topological order for e.g. 2D toric code, were obtained in
\cite{Chamon2007,Iblisdir2008,Nussinov2008}, using in \cite{Iblisdir2008} an
interesting finite-temperature extension of the topological
entanglement entropy.
Remarkably, these limitations can be overcome by
including repulsive long-range interactions with bounded strength. Such extensions
of the 2D toric code were proposed in \cite{Chesi2009} and are characterized
by a diverging relaxation time in the thermodynamic limit. Since the requirement of a
macroscopic energy barrier between logical states \cite{Bravyi2008,Kay2008} is violated in these models,
the increase of the lifetime with the system size is only polynomial.
However, the scaling power is very sensitive to the physical features of the
thermal bath and becomes especially favorable for super-ohmic reservoirs.
Such properties needed to be established in \cite{Chesi2009} by the explicit analysis of the
non-equilibrium time evolution, instead of being addressed via a suitable equilibrium quantity
as in the present work.

In \cite{Nussinov2008} a thermodynamic criterion was presented for
the existence of topological order at finite temperature. There, it
was discussed whether the thermal expectation value of logical qubit
operators could serve as a stability criterion for a quantum memory
against thermal fluctuations. Specifically, following the reasoning
used in the discussion of spontaneous symmetry breaking, a small
perturbation (external field) is applied to the system which breaks
explicitly the symmetry of the Hamiltonian and the state of the
system. Then, the thermodynamic limit is taken before the external
field  is taken to zero. If the expectation values of the logical
operators vanish in this order of limits, then, according to the
argument given in Ref.~\cite{Nussinov2008}, the information in the
quantum memory will be lost after a finite, size-independent
relaxation time at any finite temperature. This concept was
demonstrated explicitly for the Kitaev model in 2D and for some
generalizations of it to higher dimensions \cite{Nussinov2008}.

In this paper we will discuss and analyze this criterion. By making
use of elementary arguments we show that the same line of reasoning
as in Ref.~\cite{Nussinov2008} allows one to go well beyond these
results: in particular, zero thermal averages for the logical
operators are obtained not only independently of any microscopic
details of the code, be it a stabilizer code or a subsystem code,
but also in any spatial dimension. We will discuss the root cause of
these problems and discuss possible ways to extend the traditional
analysis of spontaneous symmetry breaking to detecting a finite
temperature quantum order.

The analysis of thermal stability of a quantum memory within the
formalism of the thermodynamics of open quantum systems was
seriously undertaken in a series of papers by Alicki and co-workers
\cite{AF:decay, AFH:qmem,AHHH:qmem}. In \cite{AFH:qmem} it was
demonstrated that for the 2D surface code model
weakly coupled to a Markovian environment, the relaxation rate of
any logical state is bounded from below by a constant independent of
system size \cite{AFH:qmem}. This result implies that increasing the
system size does not increase the lifetime (stability) of the
memory, but that the relaxation rate is an intrinsic feature of the
model. In \cite{AHHH:qmem} the authors considered the 4D Kitaev
model and explicitly proved that the relaxation times were
exponentially increasing with system size, hence confirming the
anticipated thermal stability in the thermodynamic limit.

In the second part of our paper (Sections~\ref{mem_ss},\ref{sec:use}) we will
present a formal analysis of the thermal stability of a quantum
memory based on subsystem (stabilizer) codes~\cite{Poulin:2005}. The
difficulty for Hamiltonian models based on subsystem codes (see
discussions in \cite{Bravyi2008}) is that the Hamiltonian is a sum
of non-commuting terms, hence spectral information for such systems
is not readily available analytically.
 Using some of the ideas developed in \cite{Nussinov2008,AF:decay,AFH:qmem,AHHH:qmem} we will construct a
simple observable whose expectation value on the thermal Gibbs state
provides an upper bound on the relaxation rate thus determining how
long quantum information can be stored in a given system.  Our
formalism will be general enough to cover both stabilizer as well as
subsystem codes.
In addition, we can use this formalism to provide a simple bound on
the memory relaxation time of stabilizer of subsystem code which is
not self-correcting, but is `protected by a gap'. In Section~\ref{sec:gapped} we prove that the
memory relaxation time scales as $n^{-1} \exp(\beta \Delta)$ where $n$ is the system size,
$\beta$ is the inverse temperature,
and $\Delta$ is the spectral gap of the memory Hamiltonian.  For sufficiently
small temperature, e.~g.~logarithmically scaling with the system size,
such models may still be of practical interest.

At the end of the paper we will show that the bound on the
relaxation rate only depends on an induced temperature-dependent
distribution associated with the Abelian stabilizer group of the
subsystem gauge group.

\section{Stabilizer and subsystem codes}
\label{sec:stabilizer_formalism}

We assume that the system chosen as the storage medium is
represented by an $n$-qubit Hilbert space ${\cal H}$. Let
$\calP_n=\la iI,X_1,Z_1,\ldots,X_n,Z_n\ra$ be the Pauli group on $n$
qubits generated by single-qubit Pauli operators and the phase
factors $\pm1 $, $\pm i$. We envision that the quantum data is
stored in the degenerate ground states of a quantum Hamiltonian
acting on the $n$ physical qubits. The Hamiltonian will be
associated with a quantum stabilizer or subsystem code.

A subsystem code is determined by its {\em gauge group} $\calG$
which can be an arbitrary subgroup of $\calG \subseteq \calP_n$. The
set of Pauli operators $P\in \calP_n$ that commute with all elements
of $\calG$ is called the {\em centralizer} of $\calG$ and is denoted
as $\calC(\calG)$. The Abelian group $\calS=\calG\cap \calC(\calG)$
is called the {\em stabilizer group} of $\calG$. If $\calG$ is Abelian, then obviously
 $\cal G=\cal S$ up to phase factors and we call $\calS$ a
stabilizer code. To preclude $\calS$ from containing non-trivial phase factors
one usually adds a requirement $-I\notin \calS$ in the case of stabilizer codes.
If $\calG$ is non-Abelian, we refer to $\calG$ as a
subsystem code.

Logical operators of a stabilizer code $\calS$  are elements of $\calC(\calS)$ which
are not in $\calS$.
One can always choose a set of logical Pauli operators
$\overline{X}_1,\overline{Z}_1, \ldots,
\overline{X}_k,\overline{Z}_k\in \calC(\calS)\backslash \calS$
obeying  the usual Pauli commutation relations:
$\overline
X_i^{2}=\overline Z_i^{2}=1$  and $\overline X_i\, \overline Z_j=(-1)^{\delta_{i,j}} \overline Z_j\, \overline X_i$.
Note that $\calC(\calS)= \langle \calS,\overline{X}_1,\overline{Z}_1, \ldots,
\overline{X}_k,\overline{Z}_k\rangle$.

The code space of a stabilizer code is defined as the common $2^k$ dimensional $+1$
eigenspace of ${\cal S}$. It can also be viewed as the ground space
of a Hamiltonian acting on $n$ qubits:
\begin{equation}\label{H_0_G_i}
H=\sum_{i=1}^m r_i S_i~.
\end{equation}
Here the $r_i$ are some real negative coupling constants and the operators
$S_i$ form an (over)complete set of generators of $\calS$.
Note that the definition of a logical operator is not
unique, since we can multiply any logical operator by an element in
$\calS$ which acts trivially on any state in the code space/ground
space. Note that the logical operators are symmetry operations of
$H$ since they commute with all elements $S_i \in {\calS}$ thus each
energy level of $H$ has a degeneracy $2^k$.
Therefore, the choice of the ground space as coding space, instead of any of
the higher energy levels, is somehow arbitrary and other forms of encoding
might be more useful. An interesting example is the thermal state encoding
which will be described in Section~\ref{sec:maps}.

{\em Bare} logical operators of a subsystem code $\calG$ are elements of
the centralizer $\calC(\calG)$ which are not in $\calG$.
One can always choose a set of bare logical Pauli operators
$\overline{X}_1,\overline{Z}_1, \ldots,
\overline{X}_k,\overline{Z}_k\in \calC(\calG)\backslash \calG$
obeying  the usual Pauli commutation relations.
Note that $\calC(\calG)= \langle \calS,\overline{X}_1,\overline{Z}_1, \ldots,
\overline{X}_k,\overline{Z}_k\rangle$.
We can
multiply such bare logical operators by elements in $\calG$ to get
so-called {\em dressed} logical operators, which act on the gauge qubits,
in addition to the logical qubits. With the group $\calG$ and its
local generators $G_i$ we can associate a Hamiltonian \be
\label{Hgauge} H=\sum_{i=1}^m r_i\, G_i, \ee where $r_i$ are some
real coefficients. Since any $G_i$ commutes with all the bare
logical operators $(\overline{X}_i,\overline{Z}_i)$, it follows that
$H$ commutes with $(\overline{X}_i,\overline{Z}_i)$. In addition,
$H$ commutes with all elements in the Abelian stabilizer group
$\calS=\calG \cap \calC(\calG)$ of $\calG$, hence $H$ is
block-diagonal in sectors labeled by the quantum numbers (syndromes)
of this stabilizer group $\calS$. Typically, ground states of $H$
are confined to a single syndrome sector.

For simplicity we will assume in the remainder of this paper that a
single qubit is encoded in the quantum memory, i.e. $k=1$.

\section{Thermal fragility?}
\label{sec:therm_frag}

To get started, let us consider the thermal fragility criterion
introduced in \cite{Nussinov2008} and apply this to general
stabilizer code Hamiltonians, Eq.~(\ref{H_0_G_i}). As in
Ref.~\cite{Nussinov2008}, we introduce $H_{\bf h}=H-{\bf h} \cdot
{\bf S}$ where the additional perturbation is a symmetry-breaking
field, designed to produce a finite expectation value of the logical
operators for the
encoded qubit. Here ${\bf S}=(\overline
X,\overline Y,\overline Z)$.

For simplicity, we will consider a perturbation ${\bf h}$ along the
$z$-direction (this can always be assumed by a suitable choice of
the logical operators), i.~e.~${\bf h}=h \hat{n_z}$ and ${\bf h}
\cdot {\bf S}=h \overline{Z}$.

We can write the degenerate eigenvectors of $H$ with energy $\epsilon_s$ as
$|s,\bf{\alpha}\rangle$ where $\alpha=\pm 1$ is the eigenvalue of
$\overline Z$ (the $H$ and the $\overline Z$ operator can be
diagonalized simultaneously).

Clearly, $H$ only acts on the $s$ quantum number (the error
syndrome, see Section \ref{sec:ec}) of the eigenfunctions $|s,\alpha
\rangle$, while the logical operators, and in particular the
perturbation $h \overline{Z}$, only acts on the $\alpha$ quantum
numbers. As a consequence, the canonical partition function $\mathcal{Z}_h={\rm
Tr}(e^{-\beta H_{\bf h}})$ at temperature $1/\beta$ factorizes
\begin{equation}\label{Z}
\mathcal{Z}_{h}=\sum_{s ,\alpha} e^{-\beta \epsilon_s}
\langle s, \alpha|e^{\;\beta h \overline{Z}}|s,\alpha\rangle
= \left( {\rm
Tr}e^{-\beta H} \right)\cosh (\beta h)  ~,
\end{equation}
where we used $\sum_\alpha \langle s, \alpha|e^{\;\beta h \overline{Z}}|s,\alpha\rangle = 2\cosh (\beta h) $,
independent of $s$.

As was shown in \cite{Nussinov2008}, we immediately see that the
average value $\langle \overline{Z} \rangle_{h}$ is independent of
the unperturbed Hamiltonian $H$, and only reflects the finite
degeneracy of the energy levels
\begin{equation}
\label{Xav} \langle \overline{Z} \rangle_{h} = \frac{\sum_{s,\alpha}
e^{-\beta \epsilon_s}\langle s,\alpha|\overline{Z} e^{\;\beta h
\overline{Z}}|s,\alpha\rangle} {\sum_{s,\alpha}e^{-\beta \epsilon_s}  \langle s, \alpha|
e^{\;\beta h \overline{Z}}|s,\alpha\rangle}=\tanh( \beta h) ~.
\end{equation}

The expectation value in Eq.~(\ref{Xav}) evidently goes to zero at
small $h$. It is clear that Eq.~(\ref{Xav}), being independent of
the form of $H$, holds also if the unperturbed Hamiltonian refers to
a macroscopic system. Therefore, this procedure yields vanishing
averages also after the thermodynamic limit is taken. If the
Hamiltonian $H$ involves $n$ physical qubits, we get
\begin{equation}
\lim_{h\to 0} \lim_{n\to \infty}\langle \overline{Z} \rangle_{h} =0.
\end{equation}
Given that this argument is independent of dimensionality, and thus
also holds for the 4D Kitaev model which is believed to be thermally
stable, the result suggests that the symmetry-breaking field used in
$H_{\bf h}$ is not strong enough to bias the thermal state $\exp(-\beta
H_{\bf h})$ towards having a non-zero logical operator expectation
value.

{\it Proof based on the Bogoliubov inequality}. We consider next an
alternative approach based on the Bogoliubov inequality
\cite{Bogoliubov62,Mermin66} and show that we reach the same
conclusion as before. This method can then be applied to subsystem
codes (see Section \ref{sec:ssc}). We start from the well-known
Bogoliubov inequality \cite{Bogoliubov62,Mermin66}
\begin{equation}
\label{bogol} \frac{\beta}{2} \langle\{ A,A^\dag \}\rangle
\langle[[C,H],C^\dag]\rangle \geq |\langle[C,A] \rangle|^2~,
\end{equation}
where $A,C$ are two arbitrary operators and $H$ is the system
Hamiltonian, with the assumption that all expectation values exist.
Here we use the convention $\{A,B\} \equiv AB+BA$. We then set
$A=\overline X$ and $C=\overline Y$. Clearly, $\{\overline
X,\overline X \}=2$ and the right-hand-side of the Bogoliubov
inequality (\ref{bogol}) gives $4 |\langle \overline Z \rangle_{h}
|^2$. Therefore we get
\begin{equation}
\beta \langle[[\overline Y,H-h \overline Z],\overline Y]\rangle_{h}
\geq 4 |\langle \overline Z \rangle_{h} |^2 ~,
\end{equation}
where the $\overline Y$ operator commutes with $H$. Using the
commutation relations for the logical operators, we obtain $4 \beta
h \langle \overline Z \rangle_{h}  \geq 4 |\langle \overline Z
\rangle_{h} |^2  \geq 0$. For $\langle \overline Z \rangle_{h}$
strictly positive (otherwise we are done) we can divide by $\langle
\overline Z \rangle_{h}$, and then, by taking the thermodynamic
limit on both sides of the resulting inequality, we eventually  get
\begin{equation}
\label{bogolZ} \lim_{n \to \infty} \langle \overline Z \rangle_{h}
\leq \beta h ~.
\end{equation}
At any finite temperature, we thus obtain that the thermal
expectation value of the logical operator vanishes when $h \to 0$.

\subsection{Subsystem codes}
\label{sec:ssc}

Let us use the Bogoliubov inequality to argue about the thermal
fragility criterion for subsystem codes. The bare logical operators
of the encoded qubit ${\bf S}=(\overline X, \overline Y, \overline
Z)$ commute with all $G_i \in \calG$, hence with $H$ in Eq.
(\ref{Hgauge}). We can choose the symmetry-breaking Hamiltonian as
\begin{equation}
H_h=H-h \overline Z G~,
\end{equation}
for some choice of $G$ which dresses the bare logical operator
$\overline{Z}$. Let us thus consider the thermal expectation value
of $\langle \overline Z G' \rangle_{h}$ where $G'$ does not need to
be the same as $G$.

We use the Bogoliubov inequality, with $A=\overline X G'$ and
$C=\overline Y$. Since $G'$ commutes with $\overline X$ and $G'
G'^\dag=1$ (valid for every Pauli operator), one obtains $
\{\overline X G',  {G'}^\dag \overline X \}=2$. This gives
\begin{equation}
\beta \langle[[\overline Y,H- h \overline Z G],\overline
Y]\rangle_{h} \geq 4 |\langle \overline Z G' \rangle_{h} |^2 ~,
\end{equation}
and since $\overline Y$ commutes with $H$ and $G$, we can easily
compute the left-hand side, to obtain
\begin{equation}
\beta h \langle \overline Z G \rangle_{h}  \geq |\langle \overline Z
G' \rangle_{h} |^2~.
\end{equation}
We now notice that $\overline Z G$ is a Pauli operator and thus has
eigenvalues $\pm 1$. Hence, the thermal expectation value on the
left side is always less then 1 independently of any details of $H$,
which gives
\begin{equation}
\lim_{n\to \infty} | \langle \overline Z G'\rangle_{h}|^2   \leq
\beta h ~,
\end{equation}
for any choice of $G$ and $G'$. Therefore, the same considerations
valid for the stabilizer codes can be repeated in this case and we
again conclude that the thermal expectation value of any logical
operator vanishes at any finite temperature for vanishing field $h$.

\section{Error correction}
\label{sec:ec}

Let us pause for a moment and discuss our somewhat naive-looking
approach. It seems that there are at least two issues at stake here.
Let us assume that by choosing the right symmetry-breaking field, we
are able to concentrate the weight of $\exp(-\beta H_{h})$ around a
logical $|\overline{0}\rangle$. Consider this $+1$ eigenstate
$|\overline{0}\rangle$ of the $\overline Z$ logical operator and a
state with a single qubit error, $E|\overline{0}\rangle$, such that
$E$ anti-commutes with $\overline Z$. Obviously, if at equilibrium
the system is in a statistical mixture of $|\overline{0}\rangle$ and
$E|\overline{0}\rangle$ with equal probability, one has $\langle
\overline Z \rangle=0$. However, the information in the memory is
still preserved as long as we correct for errors such as $E$ when we
determine what logical state has been stored. For a generic
stabilizer code, the probability of the $E|\overline{0}\rangle$
states is small at low temperature (below the gap), but the
statistical weight of all {\em correctable} errors might be very
large in the thermodynamic limit. Therefore, $\langle \overline Z
\rangle$ does not represent a meaningful stable order parameter for
this problem; the value of $\overline{Z}$ has to be modified
depending on the error syndrome. For this reason, the authors of
Ref.~\cite{AHHH:qmem} consider so-called {\em error-corrected
logical operators} \footnote{In \cite{AHHH:qmem} these are called
dressed logical operators, but we prefer to reserve the notion of
`dressing' for the multiplication of bare logical operators with
elements of the gauge group $\calG$.}. Let us properly define these
for stabilizer and subsystem codes.

For stabilizer codes, error correction consists of measuring the
$\pm 1$ eigenvalues of the stabilizer generators; these sets of
eigenvalues form the error syndrome. The error syndrome is used as
input to a classical decoding algorithm which determines which
errors have most likely taken place. For subsystem codes, error
correction may proceed by measuring the eigenvalues of the local
generators $G_i$. Since the operators $G_i$ do not commute, these
eigenvalues cannot be simultaneously measured, nonetheless these
(random) values of the generators of $\calG$ will fix the
eigenvalues of the stabilizer group $\calS$. These eigenvalues of
the stabilizer group $\calS$ again form the error syndrome.

More precisely, any error $E\in \calP_n$ determines a {\em syndrome}
$s_E\, : \, \calS \to \ZZ_2$ such that
\[
E Q=(-1)^{s_E(Q)}\, QE  \quad \mbox{for all $Q\in \calS$}.
\]
We can assume that there is some deterministic decoding algorithm
which assigns a correcting Pauli operator $C(s)\in \calP_n$ to every
syndrome $s$. An error $E\in \calP_n$ is {\em correctable} iff
$C(s_E)$ coincides with $E$ up to a gauge operator, that is,
$EC(s_E)\in \calG$.

We can define a subspace projector $P_s$ associated with every
syndrome (quantum number) $s$. Let $P_0$ be the projector onto the
$\calS$-invariant code space in which $S_i |\psi\ra=|\psi\ra$ for
all $i=1,\ldots,p$.
(By abuse of notations let us assume from now on that $-I\notin \calS$.)
For any syndrome $s$ we can define $P_s= E P_0
E^\dag$ where $E\in \calP_n$ is any error with syndrome $s$ (note
that the projector $P_s$ does not depend on the choice of such $E$).
Clearly $\sum_s P_s =I$.
 We define an \emph{error-correcting
transformation} for observables on $\calH$ as \be
\Phi_{ec}(O)=\sum_s P_s\, C(s)^\dag\, O\,   C(s) P_s. \ee
Note that $\Phi_{ec}(I)=I$, so the adjoint transformation $\Phi_{ec}^*$ acting on states
is a
trace-preserving completely-positive (TPCP) map.
Following~\cite{AHHH:qmem} we can define the {\em error-corrected
logical operators} as
 \be
\overline{Z}_{ec}=\Phi_{ec}(\overline{Z}), \quad
\overline{X}_{ec}=\Phi_{ec}(\overline{X}). \ee for a pair
of bare anti-commuting logical operators
$(\overline{X},\overline{Z})$.  Note that
$\overline{Z}_{ec},\overline{X}_{ec}$ are not necessarily Pauli
operators. However, it is not hard to show that the error-corrected
logical operators obey the relations
$\overline{Z}_{ec}^2=\overline{X}^2_{ec}=I$ and
$\overline{Z}_{ec}\overline{X}_{ec}=-\overline{X}_{ec}\overline{Z}_{ec}$.
We can understand this by defining coefficients
$\lambda_{z}(s),\lambda_x(s)\in \{+1,-1\}$ such that \be
\label{lambdas} C(s)^\dag \, \overline{X} \, C(s) =\lambda_x(s)\,
\overline{X}, \quad C(s)^\dag \, \overline{Z}\, C(s) =\lambda_z(s)\,
\overline{Z}. \ee Any syndrome projector $P_s$ belongs to the
algebra generated by $\calS$ and thus commutes with
$\overline{Z},\overline{X}$. It follows that
 \bea
 \label{dressed}
\overline{Z}_{ec}&=&\overline{Z} D_z, \quad  \mbox{where} \quad  D_z=\sum_s
\lambda_z(s)\, P_s, \\
\overline{X}_{ec}&=&\overline{X} D_x, \quad  \mbox{where}\quad
D_x=\sum_s \lambda_x(s)\, P_s.\nn
\eea
The commutation relations for $\overline{Z}_{ec},\overline{X}_{ec}$ follow
directly from Eq.~(\ref{dressed}). Note also that the
error-corrected logical operators commute with all elements in
$\calG$.

We can immediately check whether the use of error-corrected logical
operators would change the analysis of the thermal expectation
values. As observable, we choose, say, $\overline{Z}_{ec}G'$ for
some $G'$ whereas for the symmetry-breaking field we choose some
$\overline{Z}_{ec}G$. Using the properties of $\overline{Z}_{ec}$
stated above, we can repeat the proof of the previous subsection to
obtain again a vanishing expectation value
\begin{equation}
\lim_{h \to 0}\lim_{n\to \infty} \langle  \overline{Z}_{ec}
\rangle_{h} =0.
\end{equation}

\section{Analogy with the 2D Ising model: choice of symmetry-breaking field}

We emphasize that the conclusions above are valid for arbitrary
dimensions of any stabilizer or subsystem code. Although (or since)
the argument is so universal it also appears to be exceedingly
oversimplified. In the previous section, we have discussed the
necessity to choose a stable logical observable which includes the
process of error correction. Let us now more closely examine the
choice for the symmetry-breaking field.

Although the thermal fragility criterion is patterned along the
lines of standard symmetry-breaking arguments, it is only so on a
formal level. It is instructive to compare the argument of
Ref.~\cite{Nussinov2008} with the standard example of
spontaneous symmetry-breaking in the 2D Ising model \cite{yang52}
(see e.~g.~\cite{condmatbook}):
\begin{equation}\label{ising}
H_b=-J \sum_{\langle i,j \rangle} Z_i Z_j - b \sum_{i \in \Lambda}
Z_i ~,
\end{equation}
where $i, j$ label the 2D sites of the full lattice $\Lambda$, the
first sum is over pairs $\langle i,j \rangle$ of nearest neighbor
sites, and an external magnetic field $b$ is included. For the 2D
Ising model one obtains at low temperature
\begin{equation}
\lim_{b\to 0} \lim_{n\to \infty} \langle Z_j \rangle_b \neq 0~,
\end{equation}
at every lattice site $j$, where the expectation value above is
taken with respect to the Hamiltonian (\ref{ising}).
This appearance of a symmetry-breaking order should be contrasted
with the lack of such order in the 1D Ising model which has $T_c=0$.

Notice that, although the 2D Ising model does not display
topological order, it does define a proper stabilizer code with
logical operators $\overline X=\prod_{i\in \Lambda}X_i$ and
$\overline Z=Z_j$, where $j$ is a fixed (arbitrary) site in the
lattice\footnote{Of course, the other two expectation values
$\langle \overline X \rangle$ and $\langle \overline Y \rangle$ are
vanishing in the appropriate thermodynamic limit. This stabilizer
code does not provide a good quantum memory since the distance of
the code is 1 independent of lattice size.}.

The arguments discussed in the previous sections consider a
perturbation $h\overline Z=h Z_j$ which leads to $\langle \overline
Z \rangle_h=\langle Z_j \rangle_h \to 0$ and does not show that the
value of the $z$-polarization is robust. In fact, the field $h$ only
acts on a single site, whereas in the standard case the symmetry
breaking field $b$ acts on all sites of the lattice $\Lambda$
simultaneously, see Eq.~(\ref{ising}). The reason for the failure of
the stability criterion appears thus to be that the chosen
symmetry-breaking perturbation is not {\em extensive}. Although for
topological memories the support of a logical operator $\overline Z$
(i.e., the number of physical spins on which the operator acts
nontrivially) becomes larger with the size of the system, the
perturbation $h \overline Z$ is bounded in norm by $h$ and becomes
irrelevant in the thermodynamic limit.

The analogy with the 2D Ising model suggests that the
symmetry-breaking field should be chosen as a sum over different
incarnations of a logical operator, i.e. we can multiply a logical
$\overline{Z}$ by elements of the stabilizer code $\calS$ and obtain
an extensive operator. It may be possible to salvage this
symmetry-breaking route to getting a quantum order parameter, but of
course any construction should ultimately be motivated
operationally. This is the reason that we now switch to explicitly
deriving a memory relaxation rate.

\section{Relaxation rate for general quantum memory Hamiltonians}
\label{mem_ss}

The goal of this section is provide a criterion for thermal
stability for a large class of quantum systems that can be described
by subsystem codes~\cite{Poulin:2005}. This is a generalization of
the work in Ref. \cite{AHHH:qmem} in which the thermal stability of
the 4D Kitaev model was analyzed by considering the dynamics of the
quantum memory in contact with a thermal bath.

Let $\calH$ be the Hilbert space describing the system chosen as a storage media
and $\calA$ be the algebra of
operators acting on $\calH$.
The following definition will play an important role in this section.

\begin{dfn}
Let $O\in \calA$ be an observable and let $P$ be a projector
onto some subspace of $\calH$ which is invariant under $O$,
that is, $PO=OP$. We shall
say that the observable $O$ is protected from a set of errors
$\calE\subset \calA$ on a subspace $P$ iff \be \label{protected}
[E,O]\, P=0\quad \mbox{for all $E\in \calE$}. \ee
\end{dfn}
(Here and below we use the notation $P$ both for a subspace
and the corresponding projector.)
Consider as example the case when $\calE$ includes all single-qubit Pauli operators.
Suppose $O\, |\psi\ra=\lambda |\psi\ra$ for some $|\psi\ra\in P$. Then Eq.~(\ref{protected}) implies that
$O E\, |\psi\ra=\lambda E\, |\psi\ra$ for all $E\in \calE$, that is, a single-qubit error cannot change the eigenvalue of
$O$ for any eigenvector that belongs to $P$.  Quantum error correcting codes provide a systematic way of constructing
observables protected from low-weight errors on a code-subspace, see below.

Suppose for simplicity that our goal is to encode a single qubit. We
shall need a pair of observables $\tilde{X}, \, \tilde{Z}\in
\calA$ obeying the canonical commutation rules of the Pauli
operators, \be \label{PauliCCR} \tilde{X}^2=I, \quad
\tilde{Z}^2=I, \quad \tilde{X}\, \tilde{Z}=-\tilde{Z}\,
\tilde{X}. \ee
In the following we shall refer to $\tilde{X}$ and $\tilde{Z}$
obeying Eq.~(\ref{PauliCCR}) as {\em Pauli-like} observables. (Note that Pauli-like observables need not to be
single-qubit Pauli operators or tensor products of Pauli operators.)

Assume that the system evolves according to a Markovian master
equation \be \label{master1} \dot{\rho}=-i[H,\rho] + \calL(\rho),
\ee where $\calL\, :\, \calA \to \calA$ is the Lindblad operator
defined by \be \label{master2}
 \calL=\sum_{a} \, \calL_{a} \quad
\calL_a(\rho)=S_a \rho S_a^\dag - \frac12\, \{ \rho, S^\dag_a S_a\}.
\ee The operators $S_a$ will be referred to as {\em quantum jump
operators}. For any Lindblad operator $\calL$, let
$\calE_\calL\subset \calA$ be the set of all quantum jump operators
involved in $\calL$. Integrating Eq.~(\ref{master1}) one arrives at
\be \label{master3} \rho(t)=\Phi_t(\rho(0)), \quad
\Phi_t=\exp{(-it[H,\cdot] + t \calL)}. \ee We shall measure the
strength of $\calL$ using the norm \be \label{Lnorm} \| \calL\|_1
=\max_{F\in \calA}\,  \| \calL(F)\|_1 \quad \mbox{subject to
$\|F\|_1\le 1$}. \ee Here the maximization is over all self-adjoint
operators $F=F^\dag$ acting on the system Hilbert space and
$\|F\|_1$ is the trace norm of $F$, i.~e.~ $\|F\|_1={\rm Tr}
\sqrt{FF^{\dagger}}$. Note that $\|F\|_1$ is distinct from the
spectral norm $\|F\|$.

The following theorem is the main result of this section.
\begin{theorem}
\label{thm:decay}
Let $\calL$ be an arbitrary Lindblad operator
with a set of quantum jump operators $\calE_\calL$
such that the Gibbs state $\rho_\beta\sim \exp{(-\beta H)}$  is the fixed point of $\calL$,
$\calL(\rho_\beta)=0$.
Suppose one can choose Pauli-like observables
$\tilde{X},\tilde{Z}\in \calA$ that are protected from the set
of errors $\calE_\calL$  on some subspace $P$.
Suppose also that $\tilde{X},\tilde{Z}$, and $P$ commute with
the system Hamiltonian $H$. Then there exist TPCP encoding and
decoding maps $\Phi_{in}\, : \, \LL(\CC^2)\to \calA$ and
 $\Phi_{out}\, : \, \calA \to \LL(\CC^2)$ such that
\be
\label{decay_bound}
\| \Phi_t\circ \Phi_{in}(\eta) - \Phi_{in}(\eta) \|_1 \le 8t\,
\|\calL\|_1\, \trace (I-P)\rho_\beta \ee
and
\be
\Phi_{out}\circ \Phi_{in}(\eta)=\eta
\ee
 for all one-qubit states
$\eta$ and for all $t\ge 0$.
\end{theorem}
Note that the right-hand side of Eq.~(\ref{decay_bound}) provides an upper bound
on the precision up to which the decoded state $\Phi_{out}\circ \Phi_t \circ \Phi_{in}(\eta)$
approximates the initial state $\eta$. Thus assuming that the system consists of $n$ qubits
and that the norm of the Lindblad operator grows at most as $poly(n)$
we can store a qubit reliably for a time of order
\be
\label{tqmem}
\tau_{\rm qmem}\sim  (poly(n) \epsilon_{\rm qmem})^{-1},
\ee
 where
\be
\label{relaxation_rate}
\epsilon_{\rm qmem} = \trace (I-P)\rho_\beta.
\ee
We shall refer to $\tau_{\rm qmem}$ as the {\em storage time} and to the quantity $\epsilon_{\rm qmem}$ as the {\em relaxation rate}.
One can envision two scenarios when the bound Eq.~(\ref{tqmem})
on the storage time can be useful: (i) the relaxation rate $\epsilon_{\rm qmem}$
is exponentially small as a function of $n$, that is, $\epsilon_{\rm qmem}\le \exp{(-n^\gamma)}$ for some
$\gamma>0$;  (ii) the relaxation rate $\epsilon_{\rm qmem}$
is only polynomially small but the degree is
sufficiently large, such that $\tau_{\rm qmem}$ grows fast with $n$.
The first scenario can be realized for systems featuring a macroscopic (growing as $n^\gamma$) energy
barrier surrounding the states orthogonal to the protected subspace $P$.
The 4D toric code model analyzed in~\cite{AHHH:qmem} provides an example of such a system.
The second scenario could be realized if the energy barrier grows only logarithmically as
a function of $n$ as in \cite{Bacon08, HCC:toricboson}. In this case the exponent is
controlled by the temperature, that is, $\epsilon_{\rm qmem}\le n^{-\gamma \beta}=e^{-\beta \gamma \log{(n)}}$ for some $\gamma>0$.
If the temperature is smaller than  a critical value, the relaxation rate $\epsilon_{\rm qmem}$
decays sufficiently fast to yield a storage time $\tau_{\rm qmem}$ increasing with $n$.
A polynomial increase of the storage time is also obtained in \cite{Chesi2009} at any temperature, from the logarithmic divergence of a self-consistent gap.
It is tempting to conjecture that such system may exist in lower spatial dimensions.

The proof of Theorem~\ref{thm:decay} involves two ingredients: (i)
constructing the encoding and decoding maps (see
Section~\ref{subs:part1}),  and (ii) proving that the encoded states
are approximate fixed points of the Lindblad operator (see
Section~\ref{subs:part2}). Our construction of encoding and decoding
maps is identical to the one used by Alicki et al.
in~\cite{AF:decay,AHHH:qmem}. It is described in
Section~\ref{subs:part1} which can be regarded as an overview of
Section~IA in~\cite{AHHH:qmem}. The second part of the proof is
presented in Section~\ref{subs:part2}. Our approach here is quite
different from the one taken in~\cite{AHHH:qmem}. It yields a much
simpler proof and requires less assumptions about the Lindblad
operator compared to~\cite{AHHH:qmem} (for instance, we don't need
the detailed balance condition).

Following~\cite{AF:decay,AFH:qmem,AHHH:qmem} we can specialize
Theorem \ref{thm:decay} to the Markovian master equation due to
Davies~\cite{Davies} which describes the dynamics induced by a weak
coupling between the system and a thermal bath. It involves a
coupling Hamiltonian
\be H_{int} =\sum_{k=1}^K  A_k \otimes  B_k, \label{Hint} \ee where
$A_k$ are some local few-qubit operators acting on the system and
the operators $B_k$ act on the bath.

It was shown by Davies~\cite{Davies} that in the weak-coupling limit
the system evolves according to the Markovian master equation
Eq.~(\ref{master1}) where the Lindblad operator is defined as \be
\label{Davies} \calL(\rho)=\sum_k \sum_\omega h(k,\omega)
\left(A_{k,\, \omega} \rho A_{k,\, \omega}^\dag - \frac12\, \{ \rho,
A_{k,\, \omega}^\dag A_{k,\, \omega}\}\right). \ee Here $A_{k,\,
\omega}$ are the Fourier components of $A_k(t)\equiv e^{iHt} A_k
e^{-iHt}$, that is,
\[
A_k(t) = \sum_\omega A_{k,\, \omega}\, e^{-i\omega t}.
\]
One can think about $A_{k,\, \omega}$ as the part of $A_k$
transferring energy $\omega$ from the system to the bath. The bath
temperature enters into the equation only through the function
$h(k,\omega)$ which has to obey the detailed balance condition,
 \be
\label{DB}
 h(k,-\omega)=e^{-\beta \omega}\, h(k,\omega).
\ee
The coefficient $h(k,\omega)$ is defined as the Fourier transform of the autocorrelation
function of $B_k$ with respect to the bath state.
 One can regard $h(k,\omega)$
as a probability (per unit of time) of quantum jumps induced by the
coupling operator $A_k$ which transfer energy $\omega$ from the system
to the thermal bath.
The detailed balance condition guarantees that
the Gibbs state $\rho_\beta$ is a fixed point of $\calL$.

It is important to discuss how the quantum jump operators
$A_{k,\, \omega}$ depend on the original coupling operators $A_k$.

For stabilizer code Hamiltonians as in Eq.~(\ref{H_0_G_i}) the
time-dependent operator $A_k(t) =\exp(iHt) A_k \exp(-iHt)$ acts only
on a few qubits since all the terms in $H$ pairwise commute and thus
$A_k(t) =\exp(iH't) A_k \exp(-iH't)$ where $H'$ includes only those
terms of $H$ that act on the same qubits as $A_k$. Note that $H'$
has only a few Bohr frequencies since it acts only on a few qubits.
It means that any quantum jump operator $A_{k,\, \omega}$ in the
 Davies master equation acts only on a few qubits and the total number
 of the quantum jump operators is roughly the same as the number of
 the coupling operators $A_k$.

This issue is more subtle for subsystem codes, since $A_k(t)$ may be
a highly non-local operator for long times $t$ and the number of
Bohr frequencies may be exponentially large. However, it is also
clear that  the non-locality of $A_k(t)$  is {\em only} due to
multiplying it with non-local elements in the gauge group $\calG$.
Hence $A_k(t)$ remains local modulo gauge group transformations.

Let us specialize the Theorem~\ref{thm:decay} to the Davies master
equation, see Eqs.~(\ref{master1},\ref{Davies}). The condition that
the observables $\tilde{X}$ and $\tilde{Z}$ are protected from all
quantum jump operators in $\calE_\calL$ might seem too demanding
since the operators $A_{k,\, \omega}$ may be highly non-local, see
the remark above. Fortunately, it is sufficient to require that
$\tilde{X}$ and $\tilde{Z}$ are protected from a set of errors
$\calE_{int}=\{A_k\}$ including all coupling operators $A_k$.
Indeed, since, by assumption, $H$ commutes with $P$ and $\tilde{X}$,
$\tilde{Z}$, the condition
 $[\tilde{X},A_k]\, P=0$
implies $[\tilde{X}, A_{k,\, \omega}]\, P=0$ for any frequency
$\omega$. (The same remark applies to $\tilde{Z}$.)

Next we need an upper bound on the norm of the Davies generator $\calL$, see Eq.~(\ref{Davies}).
\begin{prop}
\label{prop:1}
Assuming that $\|A_k\|\le 1$ for all $k,\omega$ one has \be
\label{Lnorm_bound} \|\calL\|_1 \le 2Kh_{max}, \ee where
$h_{max}=\max_{k,\omega} |h(k,\omega)|$ and
 $K$ is the total number of terms in the interaction Hamiltonian Eq.~(\ref{Hint}).
\end{prop}
\noindent {\bf Proof.}
Indeed, let $F=F^\dag$ be an operator such that $\|F\|_1\le 1$ and $\|\calL\|_1=\|\calL(F)\|_1$,
see Eq.~(\ref{Lnorm}).
 Fix some $k$ and let $A\equiv A_k$,
$A_\omega\equiv A_{k,\, \omega}$, and $h(\omega)\equiv h(k,\omega)$.
Let us bound the trace norm of a single term
\be
\calL_k(F)\equiv \sum_\omega h(\omega)\, A_\omega F A_\omega^\dag - \frac{h(\omega)}2\, \{ A_\omega^\dag A_\omega,F\}.
\ee
 Note that \be
 \label{average1}
  \sum_\omega
A_\omega^\dag A_\omega = \lim_{T\to \infty} \frac1{2T} \int_{-T}^T
dt \, A(t)^\dag A(t)
\ee
 and
  \be
  \label{average2}
   \sum_\omega A_\omega FA_\omega^\dag
= \lim_{T\to \infty} \frac1{2T} \int_{-T}^T dt\,  A(t) F A(t)^\dag.
\ee
Using the bound $\| AB\|_1 \le \|A\|\cdot \|B\|_1$ valid for any
operators $A,B$  we get
 \bea
 \label{bound_1st_term}
 \| \sum_\omega (1/2) h(\omega)\,  \{A_\omega^\dag A_\omega, F\} \|_1 &\le&
\| \sum_\omega h(\omega)\,  A_\omega^\dag A_\omega\| \cdot \|F\|_1   \\
&\le&  h_{max} \, \| \sum_\omega A_\omega^\dag A_\omega\| \cdot \|F\|_1
\le  h_{max} \|F\|_1.  \nn
\eea
Here the second line used Eq.~(\ref{average1}), convexity of the norm, and the fact that
$\|A(t)\|=\|A\|\le 1$.

Let $F=F_+-F_-$ be the decomposition of $F$ into positive and negative parts, that is,
$F_\pm \ge 0$ and $\|F\|_1=\trace{F_-}+\trace{F_+}=\|F_-\|_1 + \|F_+\|_1$.
Then
\bea
\label{bound_2nd_term}
\| \sum_\omega h(\omega) A_\omega F A_\omega^\dag \|_1 &\le&
  \| \sum_\omega h(\omega) A_\omega F_- A_\omega^\dag \|_1+
\| \sum_\omega h(\omega) A_\omega F_+ A_\omega^\dag \|_1  \\
&\le&  h_{max} \| \sum_\omega A_\omega F_- A_\omega^\dag \|_1+
h_{max} \| \sum_\omega A_\omega F_+ A_\omega^\dag \|_1\nn  \\
&\le&  h_{max} \|F_-\|_1+ h_{max} \|F_+\|_1 = h_{max} \|F\|_1. \nn
\eea
Here the last line used Eq.~(\ref{average2}), convexity of the norm, and inequality
$\|A(t) F_\pm A(t)^\dag \|_1 \le \|F_\pm \|_1$.
 Combining Eqs.~(\ref{bound_1st_term},\ref{bound_2nd_term})
 we arrive to $\| \calL_k(F)\|_1 \le 2h_{max} \|F\|_1$ which leads to Eq.~(\ref{Lnorm_bound}).
\begin{flushright}
$\Box$
\end{flushright}

To conclude, Theorem~\ref{thm:decay} can be specialized to the
Davies master equation as follows. Suppose the system interacts with
a thermal bath at the inverse temperature $\beta$ via a Hamiltonian
$H_{int}=\sum_{k=1}^K A_k\otimes B_k$, where $\|A_k\|\le 1$ and
$B_k$ are normalized via the condition $h(k,\omega)\le h_{max}$.
Suppose one can choose Pauli-like observables $\tilde{X},\tilde{Z}$
that are protected from any coupling operator $A_k$ on some subspace
$P$. Suppose that $\tilde{X},\tilde{Z}$, and $P$ commute with the
system Hamiltonian $H$. Then Theorem~\ref{thm:decay} implies that a
qubit can be stored in the system reliably for a time $\tau_{\rm qmem}\sim (K
h_{max} \epsilon_{\rm qmem} )^{-1}$, where \be \epsilon_{\rm qmem}
\equiv \trace (I-P)\rho_\beta. \label{eq:qmem} \ee Note that $K$
will be $O(n)$ for local couplings $A_k$. We will discuss how to
evaluate the relaxation rate $\epsilon_{\rm qmem}$ in more detail in Section
\ref{sec:use}.

\subsection{Proof of Theorem~\ref{thm:decay}: part I}
\label{sec:maps}

\label{subs:part1} Let us start from defining the encoding and
decoding maps $\Phi_{in}$ and $\Phi_{out}$.  Let $\calA_Q\subseteq
\calA$ be the algebra generated by $I$, $\tilde{X}$, $\tilde{Z}$,
and $\tilde{Y}\equiv i\tilde{X}\tilde{Z}$. For any algebra $\calA$
let us define the {\em center} of $\calA$ as
\[
\calZ(\calA)=\{ A\in \calA\, : \, AB=BA\quad \mbox{for all $B\in
\calA$}\}.
\]
Clearly $\calZ(\calA_Q)=\CC\cdot I$, that is, $\calA_Q$ has trivial
center.

For any finite-dimensional Hilbert space let $\LL(\calH)$ be the
algebra of linear operators acting on $\calH$. We shall use the
following fact (see for instance Theorem~5 in~\cite{KLV:2000}, or a
book~\cite{Takesaki}):

\noindent {\bf Fact~1:} {\it Let $\calA_Q\subseteq \LL(\calH)$ be any
algebra such that (i) $\calA_Q$ contains the identity operator; (ii)
$\calA_Q$ is closed under hermitian conjugation; (iii) $\calA_Q$ has
a trivial center. Then there exists a (virtual) tensor product
structure $\calH=\calH_Q\otimes \calH_A$ such that \be
\calA_Q=\LL(\calH_Q)\otimes I_A. \ee}

It implies that there is a decomposition $\calH=\calH_Q \otimes
\calH_A$ such that $\calH_Q$ describes a qubit $Q$ and the operators
$\tilde{X},\tilde{Y},\tilde{Z}$ are the Pauli operators acting on
$\calH_Q$, that is, \be \label{Pauli_SA} \tilde{X}=X_Q\otimes I_A,
\quad \tilde{Y}=Y_Q\otimes I_A, \quad \tilde{Z}=Z_Q\otimes I_A. \ee
By assumption, the system's Hamiltonian $H$ commutes with
$\tilde{X},\tilde{Y},\tilde{Z}$. Therefore $H$ acts trivially on
$\calH_Q$ and thus there exists $H_A\in \LL(\calH_A)$ such that \be
\label{Hamiltonian_SA} H=I_Q\otimes H_A. \ee Note that
$\trace{\exp{(-\beta\, H)}}=2\trace{\exp{(-\beta\, H_A)}}$. Thus the
Gibbs state $\rho_\beta$ can be written as \be \label{Gibbs_SA}
\rho_{\beta} = \frac12\, I_Q \otimes \eta_A, \quad
\eta_A=\frac{\exp{(-\beta\, H_A)}}{\trace{\exp{(-\beta\, H_A)}}}.
\ee Define the encoding map $\Phi_{in}\, : \, \LL(\CC^2) \to \calA$
as \be \label{encoding} \Phi_{in}(\eta)=\eta \otimes \eta_A. \ee
Using Eqs.~(\ref{Pauli_SA},\ref{Hamiltonian_SA},\ref{Gibbs_SA}) one
gets \be \label{encoding1} \Phi_{in}(I)=2\rho_\beta, \quad
\Phi_{in}(Q)=2\tilde{Q} \rho_\beta=2\rho_\beta \tilde{Q} \quad
\mbox{for any $Q\in\{X,Y,Z\}$}. \ee Define the decoding map
$\Phi_{out}\, :\, \calA\to \LL(\CC^2)$ formally as the partial trace
over the subsystem $\calH_A$, \be \label{decoding}
\Phi_{out}(\rho)=\trace_A \rho. \ee Clearly, $\Phi_{out}\circ
\Phi_{in}$ is the identity map.

To demonstrate this formalism, let us explain how the encoding map
$\Phi_{in}$ is constructed for the special case of stabilizer
(subsystem) codes. Imagine that one needs to store a single qubit
state $\eta=\frac{1}{2}(I+{\bf \gamma} \cdot {\bf S})$ with ${\bf
S}=(X,Y,Z)$. We encode into the thermal state
$\Phi_{in}(\eta)=2\rho_{\beta} \eta_{ec}$ where
$\eta_{ec}=\frac12(I+{\bf \gamma} \cdot {\bf S}_{ec})$ with the
error-corrected logical operators ${\bf
S}_{ec}=(\overline{X}_{ec},\overline{Y}_{ec},\overline{Z}_{ec})$.
Note that $\eta_{ec}$ commutes with $\rho_{\beta}$.

The central idea underlying the encoding into the thermal state is
that $\Phi_{in}(\eta)$ is the same as the stationary state
$\rho_\beta$ satisfying $\Phi_t(\rho_\beta)=\rho_\beta$ on
$\calH_A$. Thus we can expect that if thermal fluctuations do not
build up to logical errors, the state $\Phi_t\circ \Phi_{in}(\eta)$
would remain close to the initial state $\Phi_{in}(\eta)$.

Note that $\Phi_{in}$ is quite different from the standard encoding into the the ground state subspace,
for which the requirement of a Hamiltonian with finite excitation gap appears most natural.
Instead, the stability criterion of Theorem~\ref{thm:decay} using
the encoding in a thermal state does not explicitly involve the
spectral gap. This is an interesting point, since it has become
clear now that the presence of a gap does not imply robustness of
topological protection. On the other hand, it might be possible to
obtain a self-correcting quantum memory for a Hamiltonian with
vanishing gap at large $n$.


\subsection{Proof of Theorem~\ref{thm:decay}: part II}
\label{subs:part2}
Let $\rho=\Phi_{in}(\eta)$ be any encoded state.
Using Eq.~(\ref{encoding1}) one check that $\rho$ can be represented
as
\be
 \rho=O\rho_\beta=\rho_\beta O, \quad O\in \calA_Q, \quad
\|O\|\le 2. \ee Consider a family of states \be \rho(t)=\Phi_t
(\rho), \quad \Phi_t=\exp{(-it[H,\cdot ]  + \calL t)}, \quad t\ge 0.
\ee Taking into account that $H$ commutes with $\rho$ we can
represent the derivative $\dot{\rho}$ as \be \dot{\rho}(s)
=\Phi_s(\calL(\rho)), \quad s\ge 0. \ee Using the fact that $\|
\Phi(A)\|_1\le \|A\|_1$ for any TPCP map $\Phi$ and any operator $A$
we get \be \| \dot{\rho}(s)\|_1 \le \| \calL(\rho)\|_1, \quad s\ge
0. \ee Therefore \be \label{dif_upper_bound} \| \rho(t)-\rho(0)\|_1
=\| \int_0^t \, ds \dot{\rho}(s) \|_1 \le t \| \calL(\rho)\|_1. \ee
Thus we have to prove an upper bound on the norm of
$\calL(\rho)=\calL(O\rho_\beta)$.
 Inserting twice the
decomposition $P+P^\perp=I$ we get
\be \calL(O\rho_\beta) =
L_1+L_2+L_3+L_4, \quad \mbox{where} \ee
\[
L_1=P\calL(OP\rho_\beta), \quad L_2=P^\perp\calL(OP\rho_\beta),
\]
and
\[
L_3=P\calL(OP^\perp\rho_\beta),\quad
L_4=P^\perp\calL(OP^\perp\rho_\beta).
\]
Using the identity $\|AB\|_1 \le \|A\| \cdot \|B\|_1$ valid for any
operators $A,B$, taking into account that $\|O\|\le 2$ and using
Eq.~(\ref{Lnorm}), one easily gets \be \label{bound_34}
\|L_3\|_1,\|L_4\|_1 \le 2\| \calL\|_1 \trace P^\perp \rho_\beta. \ee
We shall bound the norm of $L_1$ and $L_2$ using the fact that
$\calL(\rho_\beta)=0$. Indeed, using the assumption that $[S_a,O]\,
P=0$ and $P\, [S_a^\dag, O]=0$ for all $a$   one can rewrite $L_1$
as \be \label{L1simp} L_1=OP\calL(P\rho_\beta) =
-OP\calL(P^\perp\rho_\beta). \ee It follows that \be \label{bound_1}
\|L_1\|_1 \le 2\| \calL\|_1 \trace P^\perp \rho_\beta. \ee Using
$P[O,S_a^\dag]=0$, $PP^\perp=0$, and $\calL(\rho_\beta)=0$ we can
rewrite $L_2$ as \be L_2=P^\perp \calL(P\rho_\beta)O=-P^\perp
\calL(P^\perp \rho_\beta)O \ee and thus \be \label{bound_2}
\|L_2\|_1 \le 2\| \calL\|_1 \trace{P^\perp \rho_\beta}. \ee
Combining Eqs.~(\ref{bound_34},\ref{bound_1},\ref{bound_2}) we
arrive at \be \| \calL(O\rho_\beta)\|_1 \le 8\| \calL\|_1 \, \trace
P^\perp \rho_\beta. \ee Plugging it into Eq.~(\ref{dif_upper_bound})
we get $\| \rho(t)-\rho\|_1 \le 8\| \calL\|_1 \, \trace
(I-P)\rho_\beta$.

\section{Relaxation rate for subsystem code Hamiltonians}
\label{sec:use}

In this section we  explain how to construct the protected Pauli-like observables and
the subspace $P$ involved in Theorem~\ref{thm:decay}
using the formalism of subsystem codes.
Let $\calG\subseteq \calP_n$ be the gauge group of some subsystem code encoding one qubit
into $n$ qubits.
Assume that the system's Hamiltonian $H$ is defined as in Eq.~(\ref{Hgauge}), so that
$H$ is a linear combination of gauge operators.
Suppose we seek protection from some set of elementary errors $\calE$.
Assume without loss of generality that all elements of $\calE$ are Pauli operators,
that is, $\calE\subset \calP_n$. For example, $\calE$ may include all Pauli operators
that appear in the decomposition of the operators $A_k$ coupling the system and the bath,
see Eq.~(\ref{Hint}). In this case any elementary error acts only on a few qubits.

 Let us start from defining a notion of {\em goodness} of syndromes
relative to the set of elementary error
$\calE$. We will say
that a syndrome $s$, see the definitions in Section \ref{sec:ec}, is
{\em good} iff \be \label{good_syndrom} C(s+s_E) E C(s)\in \calG
\quad \mbox{for all $E\in \calE$}. \ee Remember that $C(s) \in
\calP_n$ is the correcting Pauli operator
for a given syndrome $s$ which
is determined by some deterministic error correction algorithm. To highlight
the intuition behind the definition of good syndromes, let us assume
that the syndrome $s$ has been caused by some pre-existing error
$E'$. If the error $E'$ is correctable then we have $C(s) E'\in
\calG$ and thus $EC(s)$ coincides with $EE'$ up to a gauge operator
in $\calG$. Note that $EE'$ has syndrome $s+s_E$.
Eq.~(\ref{good_syndrom}) says that $C(s+s_E)EE'\in \calG$, that is,
the error $EE'$ is also correctable for all elementary errors $E\in
\calE$.

We would like to point out that in the theory of quantum
fault-tolerance and error correction, very similar notions are used
to determine the correctness of an encoded logical gate (called a
rectangle), see e.g. in the discussion at the bottom of page 12 in
Ref.~\cite{CDT:ft}. The correctness of the encoded logical gate
depends on its incoming pre-existing syndrome in combination with
new errors which occur during the execution of the encoded gate. In
our definition of goodness, no gate happens, but we allow for any
elementary error $E$ and determine whether the pre-existing syndrome
in combination with the new error leads to making a good inference
about the total error.

Thus one can keep adding more and more elementary errors as long as
the observed syndromes are good. On the other hand, if the observed
syndrome becomes {\em bad} (that is, not good), it means that one
has already reached the limits of the error correcting capabilities
of the code and the next elementary error can potentially destroy
the encoded information. In this case the operator $C(s+s_E) E
C(s)\in \calC(\calS)\backslash \calG$ becomes a non-trivial logical
operator.

Now, in order to apply Theorem \ref{thm:decay}, we pick some
bare logical Pauli operators $\overline{X}, \overline{Z} \in \calC(\calG)\backslash \calG$,
see Section~\ref{sec:stabilizer_formalism},
and choose the Pauli-like observables $\tilde{X}$ and $\tilde{Z}$
as the error-corrected logical
operators $\overline{X}_{ec}$ and $\overline{Z}_{ec}$ defined in
Section~\ref{sec:ec}.
Below we shall prove
that  $\overline{X}_{ec}$ and $\overline{Z}_{ec}$  are protected from the
set of elementary errors $\calE$ on the subspace $P$ spanned by good syndromes, that is,
\be
 P=\sum_{{\rm good} \; s} P_s.
 \ee
Note that by construction $\overline{X}_{ec}$, $\overline{Z}_{ec}$, and $P$ commute with any Hamiltonian
$H$ made up from gauge operators, see Section~\ref{sec:ec},
while $P$ is an invariant subspace of $\overline{X}_{ec}$ and $\overline{Z}_{ec}$, see Section~\ref{sec:ec},
so all the conditions of Theorem~\ref{thm:decay} are met.

\begin{lemma}
The observable $\overline{Z}_{ec}$ is protected from all elementary
errors on the subspace spanned by good syndromes.
\end{lemma}
{\bf Proof.}
Indeed, let $s$ be any good syndrome and $E\in \calE$ be any
elementary error.
It suffices to prove that
\be
\label{aux1}
[E,\overline{Z}_{ec}]\, P_s=0.
\ee
Let $t=s+s_E$ and $A=EC(s)C(t)$. Goodness of $s$ implies that $A\in
\calG$. Since we have chosen $\overline{Z}\in \calC(\calG)$,  it
implies $A\overline{Z} = \overline{Z} A$. Taking into account
Eq.~(\ref{lambdas}) we get \be \label{comm_aux1} \overline{Z} E = E
\overline{Z} \lambda_z(s) \lambda_z(t). \ee
Thus  we see that
\[
[E,\overline{Z}_{ec}]\, P_s = P_t [E,\overline{Z}_{ec}] \, P_s= P_t
\left( \lambda_z(s) E \overline{Z} - \lambda_z(t) \overline{Z} E
\right)\, P_s =0
\]
where we have used Eqs.~(\ref{lambdas}),(\ref{dressed}) and
(\ref{comm_aux1}).

Recall that the syndrome subspaces $P_s$
are well-defined only if one fixes the subspace $P_0$ associated with the trivial syndrome.
A natural choice of $P_0$ is dictated by the ground state of the system's Hamiltonian $H=\sum_{i} r_i G_i$.
Typically the degeneracy of ground subspace of $H$ is the minimal degeneracy consistent
with the symmetry of $H$.  In our case the non-abelian symmetries of $H$ include the bare logical operators
$\overline{X}$ and $\overline{Z}$, so we should expect the ground state to have degeneracy $2$.
In this case the ground state determines a particular
set of quantum numbers (irreducible representation) of the stabilizer group
$\calS=\calG\cap \calC(\calG)$.  We can choose the trivial syndrome subspace $P_0$ as the subspace
spanned by all states that have the same quantum numbers as the ground state.
Equivalently, $P_0$ includes all states that can be obtained from the ground state by applying gauge operators
and logical operators.

The question of whether a
particular family of subsystem codes is suitable for building a good
quantum memory can now be reduced to bounding the relaxation rate
defined in Eq.~(\ref{relaxation_rate}):
 \be
 \label{subsystem_rate}
 \epsilon_{\rm
qmem}=1-\trace{\rho_\beta\, P}=\frac{\sum_{\rm bad\; s}
\calZ_s}{\sum_s \calZ_s}, \quad \mbox{where} \quad
\calZ_s=\trace{P_s\, \exp{(-\beta H)}}, \ee as a function of $n$.
Recall that all terms in $H$ commute with $\calS$ so that all syndrome subspaces
$P_s$ are invariant under $H$.
It is clearly desirable to have a Gibbs state $\exp(-\beta H)$ with
support mostly concentrated on the good syndromes. Note that
low-weight correctable errors which {\em in addition remain
correctable} if any single additional error occurs, will have
syndromes which are good. Hence the Gibbs state should be
concentrated on the ground space and excited states
that can be created from the ground state
 by  these correctable errors.
This type of property is what has been
shown for the 4D Kitaev model in~\cite{AHHH:qmem}.
Bounding the relaxation rate for subsystem codes
 is a hard task, since it
depends on the full partition function of the model.
The following simple observations might be helpful
for obtaining upper bound on $\epsilon_{\rm qmem}$.

Consider the partition function $\calZ_s$ associated with some
syndrome-sector $P_s$.
Let $E\in \calP_n$ be some error with the syndrome $s$.
Note that $P_s=EP_0E$. It implies
\bea
\calZ_s&=&
\trace{ E P_0 E \exp{(-\beta  H)}} =\trace{ P_0 \exp{(-\beta E HE)}} \nn \\
&=& \trace{ P_0 \exp{(-\beta H + \beta H_E)}}\le
\trace{P_0  \exp{(-\beta H)} \, \exp{(\beta H_E)}},\label{GT}
\eea
where
\be
H_E=2\sum_{i\, : \, G_i E=-EG_i}\, r_i G_i
\ee
is a sum of all terms in $H$ that anticommute with the error $E$.
The second line in Eq.~(\ref{GT}) follows from the Golden-Thompson inequality.
It implies
\be
\label{GTbound}
\epsilon_{\rm qmem}\le \sum_{{\rm bad}\; s} \la P_0\, \exp{  (\beta H_{E(s)})} \ra_\beta
\ee
where $\la \cdot \ra_\beta$ is the average over the thermal Gibbs state
and $E(s)$ is some fixed error causing syndrome $s$.


Let us now express our intuitive understanding of under what
circumstances the bound in Eq.~(\ref{GTbound}) could give rise to
self-correction. For a self-correcting quantum memory, we expect
that bad syndromes correspond to errors which anti-commute with a
macroscopic number of terms in $H$, hence $H_E$ is a sum over a
macroscopic number of terms, say $l(n)$. This type of requirement
has been expressed in Ref.~\cite{Bravyi2008}. If this requirement is
fulfilled, one can imagine that in `sufficiently high dimensions',
it is possible to use a mean-field approximation and approximate
$\la P_0 \exp{ (\beta H_{E(s)})} \ra_\beta$ by $\exp{(\beta  \la P_0
H_{E(s)} \ra_{\beta})}$. It is of course important that fluctuations
around such mean-field approximation die off sufficiently fast. Now,
if one can upper bound each individual term in $\la P_0 H_{E(s)}
\ra_{\beta}$ by some constant $-c$, then, because there are a
macroscopic number of terms in $H_{E(s)}$, the factor $\la P_0
\exp{(\beta H_{E(s)})} \ra_\beta$ would scale like $\exp{(-\beta
O(l(n)))}$. The sum of the bad syndromes will multiply this
exponential decay by some factor upper bounded by $2^{{\rm
rank}({\cal S})}$ where ${\rm rank}(\calS)$ is the minimal
generating set of $\calS$. If $l(n)$ grows at least as fast with $n$
as $\rank({\calS})$, then $\epsilon_{{\rm qmem}}$ would be
exponentially decaying. It is clear that many things have to 'go
right' in order for self-correction to be feasible, in particular it
is not clear whether three-dimensions would be sufficient for
mean-field type approximations with sufficiently small corrections.

Another simple observation shows that one may be able to make
headway in computing the partition functions $\calZ_s$ by making use
of the underlying symmetry. Recall that $P_s$ is the projector
associated with some irreducible representation of the stabilizer
group $\calS=\calG\cap \calC(\calG)$. Therefore we can write $P_s$
as a linear combination of elements of $\calS$, \be P_s=\sum_{Q\in
\calS} \sigma_s(Q)\, Q, \quad \sigma_s(Q) \in \{ +1,-1\}. \ee
Expanding the exponent $\exp{(-\beta H)}$ in powers of $\beta^k$ we
will only get non-zero contributions to $\calZ_s$ from terms in
$H^k$ that are elements of the stabilizer group $\calS$. We will
leave a detailed analysis of the memory relaxation rate for a
particular subsystem code to a future paper.

\section{Relaxation Rate For Gapped Hamiltonians}
\label{sec:gapped}

It is worth emphasizing that the lower bound on the storage time
obtained in Sections~\ref{mem_ss},\ref{sec:use} applies to both
gapped and gapless memory Hamiltonians. It is natural to ask whether
a stronger bound can be obtained if the memory Hamiltonian $H$ has a
constant spectral gap $\Delta$ separating the ground state from
excited states while the bath temperature $T$ is small compared to
$\Delta$. In this low-temperature regime the rate of all processes
exciting the system from the ground state is suppressed by the
Boltzmann factor $e^{-\beta \Delta}$ and thus one should expect that
the storage time scales as \be \label{gapped}
\frac1{\tau_{\mathrm{qmem}}} \le O(n) \, e^{-\beta \Delta}. \ee
Below we prove that this is indeed the case assuming that the
interaction with the thermal bath can be described by the Davies
equation, see Eqs.~(\ref{master1},\ref{Davies}). In contrast to our
main result (see Theorem~\ref{thm:decay}), the proof of
Eq.~(\ref{gapped})  will rely on the detailed balance condition
Eq.~(\ref{DB}). Our analysis will use the encoding and decoding maps
$\Phi_{in}$, $\Phi_{out}$ defined in
 Section~\ref{sec:maps} where we set $\beta=\infty$. In other words, we encode information into the ground state of
$H$ rather than the thermal state. Accordingly, any encoded state
$\rho=\Phi_{in}(\eta)$ can be represented as \be \rho=O\rho_{\infty}
=\rho_{\infty} O, \quad O\in \calA_Q, \quad \|O\|\le 2. \ee The
observable $O$ must be protected from all coupling operators $A_k$,
see Eq.~(\ref{Hint}), on the ground-state subspace $P_0$, that is,
one must have $[O,A_k]\, P_0=0$ for all $k$. As was mentioned in
Section~\ref{mem_ss}, this is equivalent to the condition \be
\label{gap1} [O,A_{k,\, \omega}]\, P_0 =0 \quad \mbox{for all $k$,
for all $\omega$}. \ee

Using Eq.~(\ref{dif_upper_bound}) it suffices to prove that
\be
\label{gap:norm}
\| \calL(\rho)\|_1 \le O(n) e^{-\beta \Delta}
\ee
where $\calL$ is the Davies generator defined in Eq.~(\ref{Davies}).
Let $A_{k,\, \omega}$ be any quantum jump operator from Eq.~(\ref{Davies}).
Since $A_{k,\, \omega}$ transfers energy $\omega$ from the system to the bath, we have
\be
A_{k,\, \omega}\,  \rho_{\infty} =0 \quad \mbox{for $\omega>0$}.
\ee
It follows that $\calL(\rho)$ contains only the terms with non-positive Bohr frequencies.
We claim that the zero-frequency term also does not contribute to $\calL(\rho)$. Indeed,
since $A_{k,0}$ commutes with $H$, we get
\be
\label{gap2}
A_{k,0} P_0 = P_0 A_{k,0}=P_0 A_{k,0} P_0.
\ee
In addition, since $A_k$ is hermitian, we have
\be
\label{gap3}
A_{k,0}^\dag=A_{k,0}.
\ee
Combining Eqs.~(\ref{gap1},\ref{gap2},\ref{gap3}) one can easily check that
\be
A_{k,0} \rho A_{k,0}^\dag - \frac12 \{ \rho,A_{k,0}^\dag A_{k,0} \} =0.
\ee
Since any negative Bohr frequency is separated from $0$ by the gap $\Delta$, we get
\be
\calL(\rho) =\sum_k  \sum_{\omega\le -\Delta}
h(k,\omega)
\left(A_{k,\, \omega} \rho A_{k,\, \omega}^\dag - \frac12\, \{ \rho,
A_{k,\, \omega}^\dag A_{k,\, \omega}\}\right).
\ee
The detailed balance condition Eq.~(\ref{DB}) implies that for any $\omega\le -\Delta$ one has
\be
\label{gap4}
|h(k,\omega)| \le e^{-\beta \Delta} \, |h(k,-\omega)| \le e^{-\beta \Delta} \, h_{max}.
\ee
Let $\calL'$ be a superoperator obtained from $\calL$ by setting $h(k,\omega)=0$
for $\omega> -\Delta$. Using Proposition~\ref{prop:1} and Eq.~(\ref{gap4})
one infers that $\|\calL'\|_1\le 2K h_{max} e^{-\beta \Delta}$ for
some $K=O(n)$. It means that \be \|\calL(\rho)\|_1
=\|\calL'(\rho)\|_1 \le \| \calL'\|_1 \le 2K h_{max} e^{-\beta
\Delta}=O(n) e^{-\beta \Delta}. \ee It proves Eq.~(\ref{gap:norm})
completing the proof of Eq.~(\ref{gapped}).

It is worth pointing out that  the condition $[O,A_k]\, P_0=0$ used in the above analysis
is satisfied whenever the coupling operators $\{ A_k\}$
are linear combinations of correctable errors with respect to the code $P_0$.
Since each coupling operator $A_k$ acts only on $O(1)$ qubits, the distance of the code $P_0$
need to be larger than some constant value depending on the locality of the coupling Hamiltonian $H_{int}$.
Thus formally the bound Eq.~(\ref{gapped}) applies even to microscopic systems that consist
only of a few qubits.  Note however that the degeneracy of the ground-space for microscopic systems is not stable
under small perturbations of the memory Hamiltonian $H$. Making the ground-space
degeneracy insensitive to perturbations requires codes with a macroscopic distance which can be achieved only
for macroscopic systems.

\section{Acknowledgements}
We thank Beat R\"othlisberger, Panos Aliferis and David DiVincenzo
for useful discussions.
SC and DL were supported by the Swiss NSF,
NCCR Nanoscience, and DARPA QUEST program under contract
number HR0011-09-1-007. SB and BMT were partially supported by the
DARPA QUEST program under contract number HR0011-09-C-0047. BMT
would like to thank the Institute for Theoretical Physics at the
University of Amsterdam for their hospitality and acknowledges
support by an ESF INSTANS exchange grant.
SB is grateful to the CWI for hospitality while this work was being
done.

\end{document}